\documentclass[fleqn,12pt,twoside]{article}
\usepackage{espcrc1}

\usepackage{graphicx}
\usepackage[figuresright]{rotating}

\newcommand{\AmS}{{\protect\the\textfont2
  A\kern-.1667em\lower.5ex\hbox{M}\kern-.125emS}}

\title{Unitarized chiral perturbation theory of hadrons}

\author{E. Oset\address {Departamento de F\'{\i}sica Te\'orica and IFIC,
        Centro Mixto Universidad de Valencia-CSIC,\\
	Ap. Correos 22085, E-46071 Valencia, Spain}%
        }
       
\begin{document}

\maketitle

\begin{abstract}
An exposition is made of recent developments using techniques of unitary chiral perturbation
theory, $U \chi P T$, which allows one to extend predictions using chiral Lagrangians to 
higher energies than ordinary chiral perturbation theory, including the region of low lying
mesonic and baryonic resonances, some of which are dynamically generated in the approach.
Results for meson meson scattering, pion and kaon form factors and meson baryon scattering
are shown. Applications are done for nuclear problems showing the results for the kaon
and eta selfenergies, phi renormalization in a nuclear medium  and $\sigma$ renormalization in
the medium, comparing results with recent experiments.  
\end{abstract}

\section{Introduction}

Chiral perturbation theory, incorporating the basic symmetries of the original QCD Lagrangian
into an effective Lagrangian which uses ordinary mesons and baryon fields as effective degrees
of freedom, has had a tremendous impact in hadronic physics at low and intermediate energies.
The theory organizes the Lagrangians in powers of the momentum of the hadrons and performs
ordinary field theoretical perturbation theory where the higher order Lagrangians provide
counterterms which regularize the theory \cite{Gasser:1984gg,Meissner:1993ah}. 
With these Lagrangians one can make predictions for meson meson interaction and
meson baryon interaction at lowest order, which reproduce the results obtained with current
algebra techniques, yet in a more elegant, systematic and technically simpler way.  The
novelties of $\chi PT$ stem from the perturbative calculations which ones performs with these
Lagrangians. The infinities appearing from the loops in second order are canceled by the
Lagrangians of next order which leave some finite counterterms. The obvious problem with 
 $\chi PT$  is its limited range of convergence. If one studies for instance s-wave meson
 meson scattering  the absolute limit for convergence is in the first pole, the one of the
 $\sigma$ meson around 500 MeV, but the lack of convergence shows already below this energy.
 More problematic is the case of the meson baryon interaction. For instance, in the study of 
 the s-wave meson baryon scattering in the strangeness $S=-1$ sector, the low energy $K^- p$
 scattering amplitude is already dominated by the $\Lambda(1405)$ resonance below threshold. 
 Ordinary  $\chi PT$ simply cannot be applied there. The region of resonances is inaccessible
 with $\chi PT$, thus putting strong limitations to the range of applicability.  Yet, the
 question arises whether  by using some suitable resummation technique one could extend this
 range of applicability, using still the content of the chiral Lagrangians. In the last years
there has been  much progress in this direction, where the consideration of the constraints of
unitarity have played a key role in the answer to the problem. In the next section we sketch 
the basic ideas about these developments.

\section{Unitarized chiral perturbation theory}
 The first steps to combine chiral perturbation theory and unitarity were done in
 \cite{Kaiser:1995eg}.  There the Lippmann Schwinger equation was used with a potential obtained
 from the lowest and higher order chiral Lagrangians. The approach was used to investigate
 $\pi N$ scattering and $\bar{K} N$ scattering around the regions of the $N^*(1525)$ and the
 $\Lambda(1405)$ resonances, respectively, and a good agreement with data was obtained using
 only a few free low energies parameters. Similarly, it was also found in
  \cite{Oller:1997ti} that the meson meson
 scattering  up to 1.2 GeV was well reproduced for s-waves using only the 
 Bethe Salpeter equation, with
 the lowest order amplitudes as kernel and a cut off of the order of 1 GeV to regularize the
 loops. The s-waves for meson baryon scattering in the $S=-1$ sector were also
 studied  in  \cite{Oset:1997it} by means of the Bethe Salpeter equation, the lowest order chiral 
 Lagrangian and a cut off, reproducing fairly well the low energy scattering properties of
 $\bar{K} N$ scattering and generating the  $\Lambda(1405)$ resonance dynamically. A further
 clarification of the issue and its extension to also p-waves in the meson meson scattering
 was done in \cite{Oller:1997ng,Oller:1998zr}, using the Inverse Amplitude Method and 
 the N/D method,
 respectively. In both cases one could obtain a good reproduction of the data and all the
 resonances up to 1.2 GeV, the $\sigma(500)$, the $f_0(980)$, the $a_0(980)$,
 the $\kappa(900)$,
 the $\rho$ and the $K^*$. 
 
  One can find a systematic and easily comprehensible derivation 
 of the  ideas of the N/D method applied for the first time to the meson baryon system in
 \cite{Oller:2000fj}, which we reproduce here below.
 One defines the transition $T-$matrix as $T_{i,j}$ between the coupled channels which couple to
 certain quantum numbers. For instance in the case of  $\bar{K} N$ scattering studied in
 \cite{Oller:2000fj} the channels with zero charge are $K^- p$, $\bar{K^0} n$, $\pi^0 \Sigma^0$,$\pi^+
 \Sigma^-$, $\pi^- \Sigma^+$, $\pi^0 \Lambda$, $\eta \Lambda$, $\eta \Sigma^0$, 
 $K^+ \Xi^-$, $K^0 \Xi^0$.
 Unitarity in coupled channels is written as
 
\begin{equation} 
Im T_{i,j} = T_{i,l} \rho_l T^*_{l,j}
\end{equation}
where $\rho_i \equiv q_i/(8\pi W)$, with $q_i$  the modulus of the c.m. 
three--momentum, and the subscripts $i$ and $j$ refer to the physical channels. 
 This equation is most efficiently written in terms of the inverse amplitude as
\begin{equation}
\label{uni}
\hbox{Im}~T^{-1}(W)_{ij}=-\rho(W)_i \delta_{ij}~,
\end{equation}
The unitarity relation in eq. (\ref{uni}) gives rise to a cut in the
$T$--matrix of partial wave amplitudes, which is usually called the unitarity or right--hand 
cut. Hence one can write down a dispersion relation for $T^{-1}(W)$ 
\begin{equation}
\label{dis}
T^{-1}(W)_{ij}=-\delta_{ij}\left\{\widetilde{a}_i(s_0)+ 
\frac{s-s_0}{\pi}\int_{s_{i}}^\infty ds' 
\frac{\rho(s')_i}{(s'-s)(s'-s_0)}\right\}+{\mathcal{T}}^{-1}(W)_{ij} ~,
\end{equation}
where $s_i$ is the value of the $s$ variable at the threshold of channel $i$ and 
${\mathcal{T}}^{-1}(W)_{ij}$ indicates other contributions coming from local and 
pole terms, as well as crossed channel dynamics but {\it without} 
right--hand cut. These extra terms
are taken directly from $\chi PT$ 
after requiring the {\em matching} of the general result to the $\chi PT$ expressions. 
Notice also that 
\begin{equation}
\label{g}
g(s)_i=\widetilde{a}_i(s_0)+ \frac{s-s_0}{\pi}\int_{s_{i}}^\infty ds' 
\frac{\rho(s')_i}{(s'-s)(s'-s_0)}
\end{equation}
is the familiar scalar loop integral
\begin{eqnarray}
\label{g2}
g(s)_i&=&\int \frac{d^4 q}{(2\pi)^4}\frac{1}{(q^2-M_i^2+i \epsilon)
((P-q)^2-m_i^2+i\epsilon)}\\
&=&\frac{1}{16 \pi^2}\left\{ a_i(\mu)+\log\frac{m_i^2}{\mu^2}+
\frac{M_i^2-m_i^2+s}{2 s}\log\frac{M_i^2}{m_i^2}+\frac{q_i}{\sqrt{s}}
\log\frac{m_i^2+M_i^2-s-2 \sqrt{s}q_i}{m_i^2+M_i^2-s+2\sqrt{s}q_i}
\right\}\nonumber  ,
\end{eqnarray}
where $M_i$ and $m_i$ are, respectively, the 
meson and baryon masses in the state $i$. In order to calculate $g(s)_i$ 
one uses the physical masses both for mesons and baryons and, hence,
 eq.(\ref{uni}) holds.  

One can further simplify the notation by employing a matrix formalism. 
Introducing the 
matrices $g(s)={\rm diag}~(g(s)_i)$, $T$ and ${\mathcal{T}}$, the latter defined in 
terms 
of the matrix elements $T_{ij}$ and ${\mathcal{T}}_{ij}$, the $T$-matrix can be written as:
\begin{equation}
\label{t}
T(W)=\left[I-{\mathcal{T}}(W)\cdot g(s) \right]^{-1}\cdot {\mathcal{T}}(W)~.
\end{equation}
which can be recast in a more familiar form as 
 \begin{equation}
\label{ta}
T(W)={\mathcal{T}}(W)+{\mathcal{T}}(W) g(s) T(W)
\end{equation}
Now imagine one is taking the lowest order chiral amplitude for the kernel as done in
\cite{Oller:2000fj}. Then the former equation is nothing but the Bethe Salpeter equation with the
kernel taken from the lowest order Lagrangian and  factorized  on  shell, the same
approach followed in \cite{Oset:1997it} where different arguments were used to justify the on shell
factorization of the kernel. Furthermore in \cite{Oller:2000fj} a simple relationship is found
between the cut off used in \cite{Oset:1997it} and the subtraction constants used in
\cite{Oller:2000fj}
\begin{equation}
\label{com}
a_i(\mu)=-2\log \left(1+\sqrt{1+\frac{m_i^2}{\mu^2}} \right)+...~,
\end{equation}
where $\mu$ plays the role of the cut off.  Then taking values of $\mu$ around 650 MeV to
1.GeV one would find subtraction constants of the order of --2, which we call of natural size.

Eq. (\ref{ta}) also serves to clarify the issue of the dynamical generation of the resonances.
According to the findings of \cite{Ecker:1989te}, the second order Lagrangian for mesons represents
the low energy limit of the exchange of vector mesons, essentially.   In that sense for the
p-waves in \cite{Oller:1998zr} the function $\mathcal{T}(W)$ contains the lowest order amplitude plus
the exchange of a bare vector meson,  and upon unitarization one can obtain a good
reproduction of the data with the right properties of mass and width of the $\rho$ and the
$K^*$ \cite{Oller:1998zr}. We call these resonances genuine, since one has to put them explicitly in the formalism
in order to obtain them, although the unitarization provides the adequate width and some extra
dressing of the resonance.  On the other hand, in the scalar sector the introduction of genuine
resonance exchange in the kernel was not needed and the unitarization of the lowest order
amplitudes gave rise to the scalar resonances \cite{Oller:1998zr}.
  In the IAM one is however making explicit use of the second order chiral 
  Lagrangians and expands $Re T^{-1}$ in powers of the momenta. In
this case the differentiation between genuine and dynamically generated resonances is not so
clear. The IAM method has also been recently used to study meson baryon interactions in
\cite{GomezNicola:1999pu}. 
  A variant of the Bethe Salpeter equation is done in
  \cite{Nieves:2000km,Nieves:2001wt}, 
  where rather than
  using the fact that one can factorize the kernel on shell introducing explicitly subtraction
  constants in $g(s)$, one assumes a certain general form for the off shell extrapolation which
  involves some free parameters and solves the Bethe Salpeter equation
  selfconsistently. 
  
  The formalism exposed here can be easily applied to the calculation of form factors, matching
  the results of the unitarized form factor to the results of one loop chiral perturbation
  theory \cite{Gasser:1984ux}.  This procedure, together with the  requirement that the 
  electromagnetic 
  pion form factor  has a peak at the $\rho$ position, and the kaon form factor at the
  $\phi$ and $\omega$ masses, determines uniquely the form factors in this
  approach, with good agreement with the data up to 1.2 GeV \cite{Oller:2000ug}.
  
  The procedure of \cite{Oller:2000fj} has been recently used to find out more 
  dynamical resonances \cite{Oset:2001cn}.
  The simple extrapolation of the approach to higher energies provides a resonance in $S=-1$
  and isospin, $I=0$, corresponding to the  $\Lambda(1670)$ resonance
 and another resonance, corresponding to $\Sigma(1620)$, which is not visible in the amplitudes
but   is found as a pole in the second Riemann sheet of the complex plane, although with a large
width.  The scheme has also been applied to the study of resonances in the $S=-2$ sector in
\cite{Ramos:2002xh}, where a $\Xi$ resonance around  1620 MeV is found.
When it comes to  compare this resonance with empirical ones one is left at the beginning with
the uncertainty to associate this resonance to the two, $I=1/2$, $\Xi$ resonances of the PDG, the 
 $\Xi(1620)$ and the   $\Xi(1690)$. 
  The study of the residues at the poles in the second Riemann sheet of the
  complex plane for the different transition
 amplitudes provides the couplings of the resonance to the different channels. 
 What one observes
 is that the couplings of the found resonance are such that the partial decay widths are totally
 incompatible with those measured for the $\Xi(1690)$ resonance, with discrepancies of the
 order of a factor 20 to 30 in all channels. This rules out completely the identification of
 the resonance found with the $\Xi(1690)$ resonance, leaving room for only the 
 $\Xi(1620)$. With
 this identification, the $\Xi(1620)$ resonance has spin and parity $1/2^-$, two magnitudes
 which were still not determined in the PDG.
 
  An interesting related work has been recently done by the Osaka-Spanish
  collaboration \cite{jido}, showing
  that one is actually obtaining two octets and a singlet of dynamically generated resonances,
 as one might expect from the SU(3) decomposition
 \begin{equation}
 8\times 8 = 1+8^s+8^a+ 10 +\bar{10}+ 27
 \end{equation}
 Indeed, if one takes the SU(3) symmetric, case when all masses of mesons on
 one side and the masses of the baryons on the other are made equal, one obtains a singlet state
 and an octet of states with zero width (since they are below the threshold for 
 the average SU(3) masses taken). When
 the symmetry is gradually broken to account for the different masses, one can
 see that the poles
 move in the complex plane and two octets appear. One finds two states with $I=0$ which move
 apart and also two states with $I=1$ which also move apart when the symmetry is
 broken. One of the $I=1$ states becomes the $\Sigma(1620)$ and the other one
 moves to lower masses close to the $\bar{K} N$ threshold and is
   too wide to have a clear repercussion in the amplitudes, but the $I=0$ states
  are narrow enough to be clearly visible in the amplitudes. In particular one
  sees that one of
  the $I=0$  states moves to the $\Lambda(1670)$ position while the other one moves to lower
  energies and mixes with the singlet state to give two resonances close to the $\Lambda(1405)$
  position. Hence what has been so far  identified as the  $\Lambda(1405)$ resonance actually
  corresponds to two poles which are close by, but which have different widths and different
  couplings to the states. 
  It might be possible to think of new reactions which give different weight to
  these two resonances. One of these 
  reactions could very well be the $K^- p \to \Lambda(1405) \gamma$ reaction,
  which was studied theoretically  in \cite{Nacher:1999ni}, where  a narrower 
  $\Lambda(1405)$ than the standard one was produced.

  In the strangeness $S=0$ sector the approach also generates the $N^*(1535)$ resonance. This
 was also  found in \cite{Kaiser:1995cy}, but it has been reviewed recently with the formalism of
 the  dispersion relation with subtraction constants in \cite{Inoue:2001ip}.

\section{Application to nuclear problems}   
 One of the important applications to nuclear physics problems has been the determination of
 the $\bar{K}$ selfenergy in a nuclear medium.  Pauli blocking corrections 
 in the intermediate nucleon states were taken into account in
 \cite{Waas:pe}, which led to a shift of the resonance and a large attractive selfenergy of the
 kaon. Subsequently a selfconsistent  calculation was done in \cite{Lutz:1997wt}, where the obtained $\bar{K}$
 selfenergy was used in the calculation, as a consequence of which the resonance moves back to
 the original position and a weaker attraction for the kaon is obtained.  Further
 developments are done in \cite{Ramos:1999ku} where, in addition, the intermediate pions and 
 baryons
 are also dressed, leading to a wider kaon spectral function and still a moderate attraction of
 the order of 40 MeV at normal nuclear matter density.  This results seemed contradictory with
 earlier expectations and fits to kaonic atoms, which demanded around 200 MeV attraction. 
 Yet, as shown in \cite{Hirenzaki:da,Baca:2000ic,Cieply:2001yg}, one can get a reasonable description 
 of the $K^-$ atoms
 with just this moderately attractive potential, which would make kaon condensation in neutron
 stars unlikely.
 
     Once one has the $\bar{K}$ potential and the  one for $K$ ( quite reliably given by the 
 $t \rho$ approximation), one can then evaluate the $\phi$ selfenergy in the nuclear medium by
 renormalizing the two kaons which come from the  $\phi$  decay into $K \bar{K}$. This was done
 in \cite{Klingl:1997kf}, subsequently in \cite{Oset:2000eg} using the improved $\bar{K}$ selfenergy of
 \cite{Ramos:1999ku}, and more recently in \cite{Cabrera:2002hc} where also the real part was calculated. 
 The results from these calculations show a very small shift of the mass and a substantial
 increase of the width which ranges from 5 to 10 times the free $\phi$ width. 
  The calculations are done for a $\phi$ at rest but one expects a similar renormalization
  for a $\phi$ moving inside the nucleus.  Although an experiment is devised in \cite{Oset:2000na} to
  measure the changes of a slow $\phi$ in nuclear $\phi$ photoproduction, we might have sooner
  results for fast moving $\phi$ in the experiment of \cite{imai} at
  Spring8/Osaka.
  
   The method of \cite{Ramos:1999ku} has also been used recently to determine the $\eta$ selfenergy
   in the nuclear medium \cite{Inoue:2002xw}.  One obtains a potential at threshold of the order 
   of (54 -i29) MeV at
   normal nuclear matter, but it also has a strong energy dependence due to the proximity of
   the $N^*(1535)$ resonance and its appreciable modification in the nuclear
   medium. One can 
   solve the Klein Gordon equation with this energy dependent potential and one finds bound
   states in medium and heavy nuclei \cite{Garcia-Recio:2002cu}, with binding
   energies ranging from  21 MeV to threshold, and half widths of the order 18
   MeV, such that  the sum of two half widths is bigger than the separation 
   between the levels. This would make the
   detection of peaks unlikely, although one could measure strength in the bound region which
   would spread up to the lowest energy plus half the width of this state, hence
   around 40 MeV
   in the bound region. The best place to eventually see these states would be
   in the region of $^{24}Mg$, where only one bound state appears with 13 MeV
   binding energy and a half width of the order of 16 MeV.
   
\begin{figure}
\center \includegraphics[angle=0,width=9.2cm]{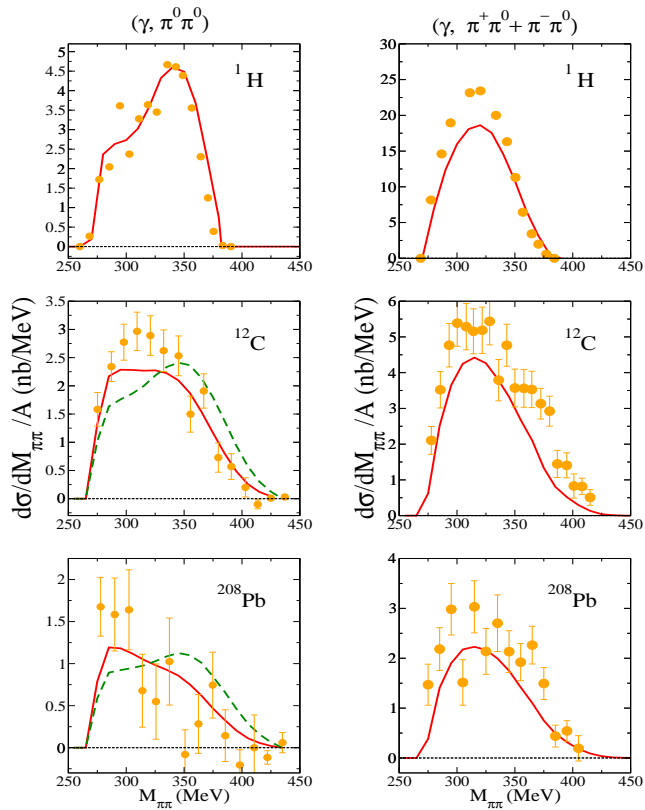}
\caption{\small{Two pion invariant mass distribution for $2\pi^0$ photoproduction 
in $^{12}C$ and $^{208}Pb$. Continuous lines: theory with $\pi \pi$ interaction
in the medium.
Dashed lines: theory with $\pi \pi$ interaction in free space. Experimental
points from \cite{Messchendorp:2002au}.}}
\label{fig:final}
\end{figure}

   Finally, let me tackle another very recent problem concerning the renormalization of the
   $\sigma$ meson in the nucleus. The $\pi \pi$ interaction in a nuclear medium has got much
   attention in the last 10 years, motivated by the original suggestion of \cite{Schuck:jn},
   where a peak appeared at nuclear matter below the two pion threshold  suggesting
   the creation of pion Cooper pairs. More refined calculations, dressing the pions in the series
   of Bethe Salpeter terms
   done in \cite{Chanfray:nn,Chiang:1997di}, indicate that the peaks do not appear but much 
   strength is
   moved to lower energies.  These results could be interpreted as the $\sigma$ mass
   getting reduced in the nuclear medium, as has been suggested in 
    \cite{Bernard:1987im,Bernard:1988sx,Hatsuda:1999kd,Jido:2001bw}, 
   and this is also the case as has been recently shown in 
   \cite{VicenteVacas:2002se} by looking at the $\sigma$ poles in the medium in
   the complex plane.
   It would be most desirable to have an experiment to test that and fortunately the experiment
  has been done recently at Mainz \cite{Messchendorp:2002au}.  It is the photoproduction of $\pi^0 \pi^0$ 
  in nuclei at small energies.  The invariant mass of the two pions has been measured and one
  observes an appreciable shift of strength of the invariant mass distribution 
  to low invariant masses as seen in fig.\ref{fig:final}.  
  
 A theoretical calculation \cite{Roca:2002vd} was done prior to the experiment, predicting this
 shift which has been confirmed by the data. Since in the chiral unitary approach the $\sigma$ is
 dynamically generated one does not have to introduce it explicitly, it simply
 comes from the consideration of the final state interaction of the two pions in s-wave and
 I=0.  Consequently with this idea, in \cite{Roca:2002vd} one takes the model
 for $(\gamma, \pi \pi)$
 of \cite{Nacher:2000eq}, which produces the two pions at tree level, and then 
 allow the pions to interact in the medium as done in \cite{Chiang:1997di}. After this is done and
 the pion absorption of the final pions from the point of production till they leave the
 nucleus is considered, following the lines of \cite{Nieves:ye}, then the
 results of fig. \ref{fig:final} are
 obtained which show a clear shift of the pion invariant mass strength to lower invariant
 masses from the proton to nuclei.  This shift is not present when the final state of the
 pions is replaced by the free one, and is also absent in the I=1 channel, both in the
 calculations and the experiment of \cite{Messchendorp:2002au}, as one can see
 in  fig. \ref{fig:final}.
   The experiment and theoretical results
 represent the first solid proof of the $\sigma$ meson renormalization in the
 nucleus, which
 should be corroborated with further calculations and experiments for other reactions. It
 also stresses the point in favor of the $\sigma$ as a dynamically generated meson by
 contrasting consequences of this hypothesis with  experiment. 
   
\section{Conclusions}  

  In the talk I have made a brief survey of the ideas about the chiral unitary approach to
study hadron dynamics at low and intermediate energies.  Then several examples of successful
application of these ideas have been shown in elementary and nuclear reactions. The
approach is powerful and represents a natural extrapolation of $\chi PT$ at higher energies.
Since it allows one to enter the regime of low lying mesonic and baryonic resonances, it
opens a broad field of possible applications, many of them already done which can not be
reported in this limited time. A review of some of the applications is done in
\cite{Oller:2000ma}. Applications to higher energies are possible with the likely introduction of
extra degrees of freedom or resonances. But the examples shown here clearly indicate that the
skillful combination of the dynamics contained  in the chiral Lagrangians and the powerful
constraints imposed by unitarity provide and ideal tool to face elementary and nuclear
reactions at intermediate energies, using a dynamics consistent with the original one of the
QCD Lagrangian and mesons and baryons as degrees of freedom, which allow an immediate
comparison with experiments  measuring directly these particles. \\

{\bf Acknowledgments}: I would like to express my gratitude to all the students and collaborators 
 who have participated in the works reported here, and who
are quoted in the references. This work is also partly
supported by DGICYT contract number BFM2000-1326 and E.U. EURIDICE network
contract no. HPRN-CT-2002.0031.

\end{document}